\documentclass[fleqn,twoside]{article}
\usepackage[headings]{espcrc2}

\readRCS
$Id: espcrc2.tex,v 1.2 2004/02/24 11:22:11 spepping Exp $
\usepackage{graphicx}
\usepackage[figuresright]{rotating}

\newcommand{\Dzb}{\overline{D^0}}
\newcommand{\barD}{\overline{D^0}}
\newcommand{\DzDzb}{D^0-\overline{D^0}}
\newcommand{\BzBzb}{B^0-\overline{B^0}}
\newcommand{\bea}{\begin{eqnarray}}
\newcommand{\eea}{\end{eqnarray}}
\newcommand{\beq}{\begin{equation}}
\newcommand{\eeq}{\end{equation}}

\newcommand{\AmS}{{\protect\the\textfont2
  A\kern-.1667em\lower.5ex\hbox{M}\kern-.125emS}}

\title{Mixing and CP-violation in charm}

\author{Alexey A. Petrov\address[MCSD]{Department of Physics and Astronomy, Wayne State University\\ 
        ~Detroit, MI 48201}%
        \thanks{Research supported by the National Science Foundation grant PHY--0244853 
	 and by the U.S. Department of Energy Contract DE-FG02-96ER41005}
       }
       
\runtitle{Mixing and CP-violation in charm}
\runauthor{A.~A.~Petrov}

\begin{document}

\begin{abstract}
The motivation most often cited in searches for $\DzDzb$ mixing and 
CP-violation in charm system lies with the possibility of 
observing a signal from new physics which dominates that from the 
Standard Model. We review recent theoretical predictions and
experimental constraints on $\DzDzb$ mixing parameters.
We also discuss the current status of searches for CP-violation in
charmed meson transitions, as well as some recent theoretical ideas. 
\vspace{1pc}
\end{abstract}

\maketitle

\section{Introduction}

Charm transitions play a unique dual role in the modern investigations of flavor physics.
It provides valuable supporting measurements for studies of CP-violation in $B$-decays, 
such as formfactors and decays constants, as well as outstanding opportunities for 
indirect searches for physics beyond the Standard Model (SM). 
It must be noted that in many dynamical models of new physics the effects of new 
particles observed in $s$, $c$, and $b$ transitions are correlated. Therefore, such 
combined studies could yield the most stringent constraints on their parameters. 
For example, loop-dominated processes such as $\DzDzb$ mixing or flavor-changing 
neutral current (FCNC) decays are influenced by the dynamical effects of 
{\it down-type particles}~\cite{Sandip}, whereas up-type particles are responsible 
for FCNC in the beauty and strange systems. Finally, from the practical point of 
view, charm physics experiments provide outstanding opportunities for studies of 
new physics because of the availability of large statistical samples of data.

The basic idea behind searches for new physics in the relatively low energy 
processes like charm transitions stems from the fact that the effects of new 
interactions in can be naturally written in terms of a series of local operators 
of increasing dimension. Together with the one loop Standard Model effects~\cite{Petrov:1997ch} 
they can contribute to $\Delta C = 1$ (decays) or $\Delta C = 2$ (mixing) FCNC transitions. 
For example, in the case of $\DzDzb$ mixing these operators generate contributions 
to the effective operators that change $D^0$ state into $\Dzb$ state,
leading to the mass eigenstates
\begin{equation} \label{definition1}
| D_{^1_2} \rangle =
p | D^0 \rangle \pm q | \bar D^0 \rangle,
\end{equation}
where the complex parameters $p$ and $q$ are obtained from diagonalizing 
the $D^0-\Dzb$ mass matrix. Note that $|p|^2 + |q|^2 = 1$. If CP-violation
in mixing is neglected, $p$ becomes equal to $q$, so $| D_{1,2} \rangle$ 
become $CP$ eigenstates, $CP | D_{\pm} \rangle = \pm | D_{\pm} \rangle$.
The mass and width splittings between these eigenstates are given by
\begin{eqnarray} \label{definition}
x \equiv \frac{m_2-m_1}{\Gamma}, ~~
y \equiv \frac{\Gamma_2 - \Gamma_1}{2 \Gamma}.
\end{eqnarray}
It is known experimentally that $\DzDzb$ mixing proceeds extremely slowly, 
which in the Standard Model is usually attributed to the absence of superheavy 
quarks destroying GIM cancellations. This situation is an exact opposite to what 
happens in the $B$ system, where $\BzBzb$ mixing measurements are used to 
constrain top quark couplings~\cite{Bphysics}.

It is instructive to see how new physics can affect charm mixing.
Since the lifetime difference $y$ is constructed from the decays of $D$ into 
physical states, it should be dominated by the Standard Model contributions, 
unless new physics significantly modifies $\Delta C=1$ interactions. On the 
contrary, the mass difference $x$ can receive contributions from all energy scales.
Thus, it is usually conjectured that new physics can significantly
modify $x$ leading to the inequality $x\gg y$~\footnote{This signal for new physics 
is lost if a relatively large $y$, of the order of a percent, 
is observed~\cite{Bergmann:2000id}.}. 

Another possible manifestation of new physics interactions in the charm
system is associated with the observation of (large) CP-violation. This 
is due to the fact that all quarks that build up the hadronic states in weak 
decays of charm mesons belong to the first two generations. Since $2\times2$ 
Cabbibo quark mixing matrix is real, no CP-violation is possible in the
dominant tree-level diagrams which describe the decay amplitudes. 
CP-violating amplitudes can be introduced in the Standard Model by including 
penguin or box operators induced by virtual $b$-quarks. However, their 
contributions are strongly suppressed by the small combination of 
CKM matrix elements $V_{cb}V^*_{ub}$. It is thus widely believed that the 
observation of (large) CP violation in charm decays or mixing would be an 
unambiguous sign for new physics. This fact makes charm decays a valuable 
tool in searching for new physics, since the statistics available in charm 
physics experiment is usually quite large.

As in B-physics, CP-violating contributions in charm can be generally 
classified by three different categories:
(I) CP violation in the decay amplitudes. This type of CP violation 
occurs when the absolute value of the decay amplitude for $D$ to decay to a 
final state $f$ ($A_f$) is different from the one of corresponding 
CP-conjugated 
amplitude (``direct CP-violation'');
(II) CP violation in $\DzDzb$ mixing matrix. This type of CP violation
is manifest when 
$R_m^2=\left|p/q\right|^2=(2 M_{12}-i \Gamma_{12})/(2 M_{12}^*-i 
\Gamma_{12}^*) \neq 1$; 
and 
(III) CP violation in the interference of decays with and without mixing.
This type of CP violation is possible for a subset of final states to which
both $D^0$ and $\Dzb$ can decay. 

For a given final state $f$, CP violating contributions can be summarized 
in the parameter 
\begin{equation}
\lambda_f = \frac{q}{p} \frac{{\overline A}_f}{A_f}=
R_m e^{i(\phi+\delta)}\left| \frac{{\overline A}_f}{A_f}\right|,
\end{equation}
where $A_f$ and ${\overline A}_f$ are the amplitudes for $D^0 \to f$ and 
$\Dzb \to f$ transitions respectively and $\delta$ is the strong phase 
difference between $A_f$ and ${\overline A}_f$. Here $\phi$ represents the
convention-independent weak phase difference between the ratio of 
decay amplitudes and the mixing matrix. Since CP-violation in the mixing matrix 
is expected to be small, we will often expand
$R_m^{\pm 2} = 1 \pm A_m$~\cite{Bergmann:2000id}.

At present, most experimental information about the $\DzDzb$ mixing parameters 
$x$ and $y$ comes from the time-dependent analyses that can roughly be divided
into two categories. First, more traditional studies look at the time
dependence of $D \to f$ decays, where $f$ is the final state that can be
used to tag the flavor of the decayed meson. The most popular is the
non-leptonic doubly Cabibbo suppressed (DCS) decay $D^0 \to K^+ \pi^-$.
Time-dependent studies allow one to separate the contribution of direct 
doubly Cabbibo suppressed transition from the one involving 
mixing $D^0 \to \Dzb \to K^+ \pi^-$,
\begin{eqnarray}\label{Kpi}
\Gamma[D^0 \to K^+ \pi^-]
=e^{-\Gamma t}|A_{K^-\pi^+}|^2 ~~~~~~~~~
\nonumber \\
\times ~\left[ R + \sqrt{R} R_m(y'\cos\phi 
-  x'\sin\phi)\Gamma t \right.
\\
\left. +~\frac{R_m^2}{4}(y^2+x^2)(\Gamma t)^2
\right],
\nonumber
\end{eqnarray}
where $\sqrt{R} e^{i \delta}$ parameterizes the ratio of DCS and Cabibbo favored 
(CF) decay amplitudes, and $x'=x\cos\delta+y\sin\delta$, 
$y'=y\cos\delta-x\sin\delta$. Since $x$ and $y$ are small, the best constraint 
comes from the linear terms in $t$ that are also {\it linear} in $x$ and $y$.
As follows from Eq.~(\ref{Kpi}) direct extraction of $x$ and $y$ is not possible due 
to unknown relative strong phase $\delta$ of DCS and CF amplitudes~\cite{Falk:1999ts}. 
This phase, however, can be measured in an experiment which produces CP-correlated pairs 
of $D$-mesons, such as CLEO-c~\cite{Gronau:2001nr}. Recent experimental constraints on
$\DzDzb$ mixing parameters from $D^0 \to K^+ \pi^-$ analyses are presented in 
Table~\ref{table:1} (presented Belle and FOCUS results assume conservation of CP).
\begin{table*}[htb]
\caption{Recent measurements of $x$ and $y$ in $D^0 \to K^+ \pi^-$}
\label{table:1}
\newcommand{\m}{\hphantom{$-$}}
\newcommand{\cc}[1]{\multicolumn{1}{c}{#1}}
\renewcommand{\tabcolsep}{2pc} 
\renewcommand{\arraystretch}{1.2} 
\begin{tabular}{@{}lll}
\hline
Experiment           & $x'^2$ (95\% CL) & $y'$ (95\% CL) \\
$~$                  & $~~(\times 10^{-3})$ & $~~(\times 10^{-3})$ \\
\hline
Belle (2004)               & $<0.81$ & $-8.2<y'<16$ \\
BaBar (2003)               & $<2.2$ & $-56<y'<39$ \\
FOCUS (2001)               & $<1.52$ & $-124<y'<-5$ \\
CLEO (2000)                & $<0.82$ & $-58<y'<10$ \\
\hline
\end{tabular}\\[2pt]
The experimental values are given in ref.~\cite{KpiExp}.
\end{table*}

Second, $D^0$ mixing parameters can be measured by comparing the lifetimes 
extracted from the analysis of $D$ decays into the CP-even and CP-odd 
final states. This study is also sensitive to a {\it linear} function of 
$y$ via
\begin{equation}
\frac{\tau(D \to K^-\pi^+)}{\tau(D \to K^+K^-)}-1=
y \cos \phi - x \sin \phi \frac{A_m}{2}.
\end{equation}
The results from various experiments can be found in ref.~\cite{Yabsley:2003rn},
with the grand average $y_{CP}=(0.9\pm0.4)\%$.

Time-integrated studies of the semileptonic transitions are sensitive
to the {\it quadratic} form $x^2+y^2$ and at the moment are not 
competitive with the analyses discussed above. 

The construction of tau-charm factories CLEO-c and 
BES-III allows for new {\it time-independent} methods that 
are sensitive to a linear function of $y$. One can use the 
fact that heavy meson pairs produced in the decays of heavy quarkonium 
resonances have the useful property that the two mesons are in the CP-correlated 
states~\cite{Atwood:2002ak}. For instance, by tagging one of the mesons as a CP 
eigenstate, a lifetime difference 
may be determined by measuring the leptonic branching ratio of the other meson.
Its semileptonic {\it width} should be independent of the CP quantum number 
since it is flavor specific, yet its {\it branching ratio} will be inversely 
proportional to the total width of that meson. Since we know whether this $D(k_2)$ state is 
tagged as a (CP-eigenstate) $D_\pm$ from the decay of $D(k_1)$ to a 
final state $S_\sigma$ of definite CP-parity $\sigma=\pm$, we can 
easily determine $y$ in terms of the semileptonic branching ratios of $D_\pm$, which
we denote $\cal{B}^{\ell}_\pm$. Neglecting small CP-violating effects,
\begin{equation} 
y=\frac{1}{4} \left(
\frac{\cal{B}^{\ell}_+(D)}{\cal{B}^{\ell}_-(D)}-
\frac{\cal{B}^{\ell}_-(D)}{\cal{B}^{\ell}_+(D)}
\right).
\end{equation}
A more sophisticated version of this formula as well as studies of 
feasibility of this method can be found in ref.~\cite{Atwood:2002ak}.

The current experimental upper bounds on $x$ and $y$ are on the order of 
a few times $10^{-2}$, and are expected to improve significantly in the coming
years.  To regard a future discovery of nonzero $x$ or $y$ as a signal for new 
physics, we would need high confidence that the Standard Model predictions lie
well below the present limits.  As was recently shown~\cite{Falk:2001hx}, 
in the Standard Model, $x$ and $y$ are generated only at second order in SU(3)$_F$ 
breaking, 
\begin{equation}
x\,,\, y \sim \sin^2\theta_C \times [SU(3) \mbox{ breaking}]^2\,,
\end{equation}
where $\theta_C$ is the Cabibbo angle.  Therefore, predicting the
Standard Model values of $x$ and $y$ depends crucially on estimating the 
size of SU(3)$_F$ breaking.  Although $y$ is expected to be determined
by the Standard Model processes, its value nevertheless affects significantly 
the sensitivity to new physics of experimental analyses of $D$ 
mixing~\cite{Bergmann:2000id}.

%
\begin{figure}[htb]
\begin{center}
\includegraphics[scale=0.32]{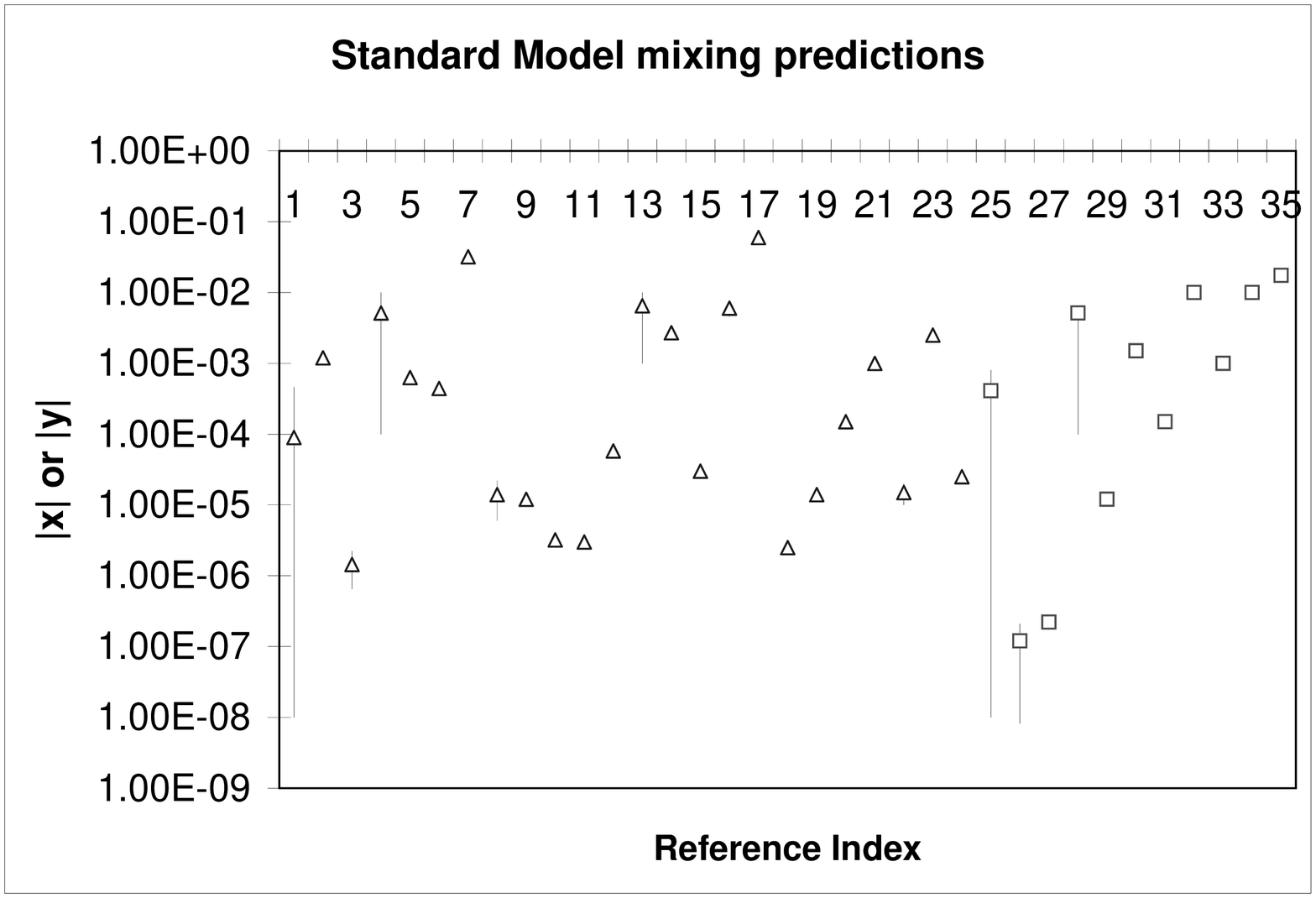}
\caption{Standard Model predictions for $|x|$ (open triangles) 
and $|y|$ (open squares). Horizontal line references are 
tabulated in~\cite{Nelson:1999fg}.}
\label{fig:DDmixSM}
\end{center}
\end{figure}
Theoretical predictions of $x$ and $y$ within and beyond
the Standard Model span several orders of magnitude~\cite{Nelson:1999fg} 
(see Fig.~\ref{fig:DDmixSM}).
Roughly, there are two approaches, neither of which give very reliable
results because $m_c$ is in some sense intermediate between heavy and
light.  The ``inclusive'' approach is based on the operator
product expansion (OPE).  In the $m_c \gg \Lambda$ limit, where
$\Lambda$ is a scale characteristic of the strong interactions, $\Delta
M$ and $\Delta\Gamma$ can be expanded in terms of matrix elements of local
operators~\cite{Inclusive}.  Such typically calculations yield $x,y < 10^{-3}$.  
The use of the OPE relies on local quark-hadron duality, 
and on $\Lambda/m_c$ being small enough to allow a truncation of the series
after the first few terms.  The charm mass may not be large enough for these 
to be good approximations, especially for nonleptonic $D$ decays.
An observation of $y$ of order $10^{-2}$ could be ascribed to a
breakdown of the OPE or of duality,  but such a large
value of $y$ is certainly not a generic prediction of OPE analyses.
The ``exclusive'' approach sums over intermediate hadronic
states, which may be modeled or fit to experimental data~\cite{Exclusive}.
Since there are cancellations between states within a given $SU(3)$
multiplet, one needs to know the contribution of each state with high 
precision. However, the $D$ is not light enough that its decays are dominated
by a few final states.  In the absence of sufficiently precise data on many decay 
rates and on strong phases, one is forced to use some assumptions. While most 
studies find $x,y < 10^{-3}$, Refs.~\cite{Exclusive} obtain $x$ and 
$y$ at the $10^{-2}$ level by arguing that SU(3)$_F$ violation is of order
unity. 
It was also shown that phase space effects alone provide enough SU(3)$_F$ 
violation to induce $x,y\sim10^{-2}$~\cite{Falk:2001hx}.
Large effects in $y$ appear for decays close to $D$ threshold, where
an analytic expansion in SU(3)$_F$ violation is no longer possible; a dispersion 
relation can then be used to show that $x$ would receive contributions of
similar order of magnitude.

The above discussion shows that, contrary to $B$ and $K$ systems, theoretical 
calculations of $x$ and $y$ are quite uncertain~\cite{Reviews}, and the values
near the current experimental bounds cannot be ruled out. Therefore, it will 
be difficult to find a clear indication of physics beyond the Standard Model 
in $\DzDzb$ mixing measurements alone. The only robust potential signal of new 
physics in charm system at this stage is observation of large violation of CP.

CP violation in $D$ decays and mixing can be searched for by a variety of 
methods. For instance, time-dependent decay widths for $D \to K \pi$ are 
sensitive to CP violation in mixing (see Eq.(\ref{Kpi})). Provided that 
the $x$ and $y$ are comparable to experimental sensitivities, a combined 
analysis of $D \to K \pi$ and $D \to KK$ can yield interesting 
constraints on CP-violating parameters~\cite{Bergmann:2000id}.

Most of the techniques that are sensitive to CP violation make use of the
decay asymmetry,
\begin{eqnarray}\label{Acp}
A_{CP}(f)=\frac{\Gamma(D \to f)-\Gamma({\overline D} \to {\overline f})}{
\Gamma(D \to f)+\Gamma({\overline D} \to {\overline f})}.
\end{eqnarray}
Most of the properties of Eq.~(\ref{Acp}), such as dependence on the
strong final state phases, are similar to the ones in B-physics~\cite{BigiSandaBook}.
Current experimental bounds from various experiments, all consistent
with zero within experimental uncertainties, can be found in~\cite{Pedrini:2000ge}.

Other interesting signals of $CP$-violation that are being 
discussed in connection with tau-charm factory measurements are
the ones that are using quantum coherence of the initial state.
An example of this type of signal is a decay $(D^0 \Dzb) \to f_1 f_2$ at 
$\psi(3770)$ with $f_1$ and $f_2$ being the different final CP-eigenstates
of the same CP-parity. This type of signals are very easy to detect 
experimentally. The corresponding CP-violating decay rate for the final states
$f_1$ and $f_2$ is
\begin{eqnarray} \label{CPrate}
\Gamma_{f_1 f_2} &=&
\frac{1}{2 R_m^2} \left[
\left(2+x^2-y^2\right) \left|\lambda_{f_1}-\lambda_{f_2}\right|^2
\right.
\nonumber \\
&+& \left .\left(x^2+y^2\right)\left|1-\lambda_{f_1} \lambda_{f_2}\right|^2
\right]~\Gamma_{f_1} \Gamma_{f_2}.
\end{eqnarray}
The result of Eq.~(\ref{CPrate}) represents a slight generalization of the formula 
given in Ref.~\cite{Bigi:1986dp}. It is clear that both terms in the numerator 
of Eq.~(\ref{CPrate}) receive contributions from CP-violation of the type I 
and III, while the second term is also sensitive to CP-violation of the
type II. Moreover, for a large set of the final states the first term would be 
additionally suppressed by SU(3)$_F$ symmetry, as for instance, 
$\lambda_{\pi\pi}=\lambda_{KK}$ in the SU(3)$_F$ symmetry limit. 
This expression is of the {\it second} order in CP-violating parameters 
(it is easy to see that in the approximation where only CP violation in the mixing 
matrix is retained, $\Gamma_{f_1 f_2} \propto \left|1-R_m^2\right|^2 \propto A_m^2$).

The existing experimental constraints~\cite{Csorna:2001ww} demonstrate that 
CP-violating parameters are quite small in the charm sector, regardless of 
whether they are produced by the Standard Model mechanisms or by some new physics 
contributions. Since the above measurements involve CP-violating decay {\it rates}, 
these observables are of {\it second order} in the small CP-violating 
parameters, a challenging measurement.

It is also easy to see that the rate asymmetries of 
Eq.~(\ref{Acp}) require tagging of the initial state with the consequent 
reduction of the dataset. In that respect, it is important to maximally exploit
the available statistics. 

It is possible to use a method that both does not require flavor or CP-tagging of the 
initial state and results in the observable that is {\it first order} 
in CP violating parameters~\cite{Petrov:2004gs}. Let's concentrate on the 
decays of $D$-mesons to final states that are common for $D^0$ and $\barD$. 
If the initial state is not tagged the quantities that one can easily measure 
are the sums 
\begin{equation} 
\Sigma_i=\Gamma_i(t)+{\overline \Gamma}_i(t) 
\end{equation}
for $i=f$ and ${\overline f}$.
A CP-odd observable which can be formed out of $\Sigma_i$ 
is the asymmetry
\begin{equation} \label{TotAsym}
A_{CP}^U (f,t) =  
\frac{\Sigma_f - \Sigma_{\overline f}}{\Sigma_f + \Sigma_{\overline f}}
\equiv \frac{N(t)}{D(t)}.
\end{equation}
We shall consider both time-dependent and time-integrated versions 
of the asymmetry (\ref{TotAsym}). Note that this asymmetry does not 
require quantum coherence of the initial state and therefore is accessible in 
any $D$-physics experiment. It is expected that the numerator and denominator 
of Eq.~(\ref{TotAsym}) would have the form,
\bea \label{NumDenum}
N(t) &=& \Sigma_f - \Sigma_{\overline f} 
= ~e^{-{\cal T}} \left[A + B {\cal T} + C {\cal T}^2 \right],
\nonumber \\
D(t) &=& ~2 e^{-{\cal T}} \left[ 
\left | A_f \right|^2 +  
\left | A_{\overline f} \right|^2 
\right],~~~~
\eea
where we neglected direct CP violation in $D(t)$.
Integrating the numerator and denominator of Eq.~(\ref{TotAsym}) over time 
yields
\beq
A_{CP}^U (f) = \frac{1}{D}\left[A + B + 2 C\right],
\eeq
where $D=\Gamma \int_0^\infty dt ~D(t)$. 

Both time-dependent and time-integrated asymmetries depend on the same parameters
$A, B$, and $C$. The result is
\bea\label{Coefficients}
A &=& \left | A_f \right|^2 -  \left | \overline A_{\overline f} \right|^2 -
\left | A_{\overline f} \right|^2 +  \left | \overline A_f \right|^2,
\nonumber \\
B &=& 
-2 y \sqrt{R} ~\left[ \sin\phi \sin \delta 
\left(
\left | \overline A_f \right|^2 + \left | A_{\overline f} \right|^2
\right) \right. \nonumber \\
&& -~ \left. \cos\phi \cos \delta
\left(
\left | \overline A_f \right|^2 - \left | A_{\overline f} \right|^2
\right)
\right],
\\
C &=& \frac{x^2}{2} A.
\nonumber
\eea
We neglect small corrections of the order of ${\cal O}(A_m x, r_f x, ...)$ and
higher. It follows that Eq.~(\ref{Coefficients}) receives contributions from both 
direct and indirect CP-violating amplitudes. Those contributions have different time
dependence and can be separated either by time-dependent analysis of Eq.~(\ref{TotAsym}) 
or by the ``designer'' choice of the final state. Note that this asymmetry is manifestly 
{\it first} order in CP-violating parameters.

In Eq.~(\ref{Coefficients}), non-zero value of the coefficient $A$ is an indication 
of direct CP violation. This term is important for singly Cabibbo suppressed (SCS) decays. 
The coefficient $B$ gives a combination of a contribution of CP violation in the 
interference of the decays with and without mixing (first term) and direct CP 
violation (second term). Those contributions can be separated by considering 
DCS decays, such as $D \to K^{(*)} \pi$ or $D \to K^{(*)} \rho$, where direct CP violation 
is not expected to enter. The coefficient $C$ represents a contribution of 
CP-violation in the decay amplitudes after mixing. It is negligibly small in the SM and 
all models of new physics constrained by the experimental data. Note that the effect of 
CP-violation in the mixing matrix on $A$, $B$, and $C$ is always subleading.

Eq.~(\ref{Coefficients}) is completely general and is true for both DCS and SCS 
transitions. Neglecting direct CP violation we obtain a much simpler expression,
\bea \label{KpiTime}
A &=& 0, \qquad C = 0, \nonumber \\
B &=& - 2 y \sin\delta \sin\phi ~\sqrt{R} ~
\left[
\left | \overline A_f \right|^2 + \left | A_{\overline f} \right|^2
\right]
\eea
For an experimentally interesting DCS decay $D^0 \to K^+ \pi^-$
this asymmetry is zero in the flavor $SU(3)_F$ symmetry limit, where
$\delta = 0$~\cite{Wolfenstein:1985ft}. Since $SU(3)_F$ is badly broken 
in $D$-decays, large values of $\sin\delta$~\cite{Falk:1999ts} are possible. 
At any rate, regardless of the theoretical estimates, this strong phase could be 
measured at CLEO-c.
It is also easy to obtain the time-integrated asymmetry for $K\pi$. Neglecting small
subleading terms of ${\cal O}(\lambda^4)$ in both numerator and 
denominator we obtain
\beq \label{KpiIntegrated}
A_{CP}^U (K\pi) = - y \sin \delta \sin \phi \sqrt{R}. 
\eeq
It is important to note that both time-dependent and time-integrated asymmetries
of Eqs.~(\ref{KpiTime}) and (\ref{KpiIntegrated}) are independent of predictions
of hadronic parameters, as both $\delta$ and $R$ are experimentally determined 
quantities and could be used for model-independent extraction of CP-violating phase 
$\phi$. Assuming $R \sim 0.4\%$ and $\delta \sim 40^o$~\cite{Falk:1999ts} and
$y \sim 1\%$ one obtains 
$\left| A_{CP}^U (K\pi) \right| \sim \left(0.04\%\right)\sin\phi$.
Thus, one possible challenge of the analysis of the asymmetry 
Eq.~(\ref{KpiIntegrated}), is that it involves a difference of two large 
rates, $\Sigma_{K^+\pi^-}$ and $\Sigma_{K^-\pi^+}$, which should be
measured with the sufficient precision to be sensitive to $A_{CP}^U$, 
a problem tackled in determinations of tagged asymmetries in $D \to K \pi$ 
transitions.

Alternatively, one can study SCS modes, where $R \sim 1$, so the 
resulting asymmetry could be ${\cal O}(1\%)\sin\phi$. However, the final states 
must be chosen such that $A_{CP}^U$ is not trivially zero. For example, 
decays of $D$ into the final states that are CP-eigenstates would result 
in zero asymmetry (as $\Gamma_f=\Gamma_{\overline f}$ for those final states) 
while decays to final states like $K^+ K^{*-}$ or $\rho^+ \pi^-$ would not.
It is also likely that this asymmetry is larger than the estimate given above 
due to contributions from direct CP-violation (see eq.~\ref{Coefficients}). 

The final state $f$ can also be a multiparticle state. In that case,
more untagged CP-violating observables could be constructed. For example,
untagged studies of Dalitz plot population asymmetries resulting from 
the enantiometric intermediate states were proposed in~\cite{Gardner:2003su} 
to study direct CP-violation in $B_d$ decays. A similar study is possible here
as well.

As any rate asymmetry, Eq.~(\ref{TotAsym}) requires either a ``symmetric'' 
production of $D^0$ and $\barD$, a condition which is automatically satisfied 
by all $p\overline p$ and $e^+ e^-$ colliders, or a correction for 
$D^0/\barD$ production asymmetry.

In summary, charm physics, and in particular studies of CP-violation,
could provide new and unique opportunities for indirect searches for new physics. 
Expected large statistical samples of charm data will allow new sensitive
measurements of charm mixing and CP-violating parameters.


\end{document}